\documentclass[12pt, a4paper]{article}

 \usepackage[bottom]{footmisc}

\usepackage{graphicx}
\usepackage{amsthm}   
\usepackage{amsmath} 
\usepackage{amssymb}  
\usepackage{mathrsfs} 
\usepackage{stmaryrd} 
\usepackage{txfonts} 

\usepackage{hyperref}  
\usepackage[capitalize]{cleveref}

\usepackage{enumerate}
\usepackage{appendix} 
\usepackage{comment}
\usepackage{notoccite}

\usepackage[all]{xy}

\newcommand{\mathsym}[1]{{}}
\newcommand{\unicode}[1]{{}}

\newcommand{\R}{{\mathbb{R}}}

\newcommand{\N}{{\mathbb{N}}}

\newcommand{\U}{{\mathcal{U}}}

\newcommand{\hatout}{\widehat{out}}

\usepackage{color}  
\RequirePackage[dvipsnames,usenames]{xcolor}

\newcommand{\dhwc}[1] {{\color{Red}{\bf{DHW COMMENT: #1}}}}

\newcommand{\prove}[1] {{\color{blue}{\bf{Prove. }}}}

\usepackage[obeyFinal]{todonotes}
\usepackage{ifthen,xspace}

\newcommand{\eq}[1]{\begin{align}#1\end{align}}

\newcommand{\be}{\begin{equation}}
\newcommand{\bel}[1]{\begin{equation}\label{#1}}
\newcommand{\qe}{\end{equation}}
\newcommand{\ee}{\end{equation}}
\newcommand{\eeq}{\end{equation}}
\newcommand{\ba}{\begin{eqnarray}}
\newcommand{\ea}{\end{eqnarray}}

\def\bal#1\eal{\begin{align}#1\end{align}}
\def\bann#1\eann{\begin{align*}#1\end{align*}}

\date{\today}                      

\begin{document}

\title{Stochastic Process Turing Machines}

\author{David H. Wolpert \\
 Santa Fe Institute, 1399 Hyde Park Road, Santa Fe, NM, 87501\\
 \emph{and} \\
 Jordan Scharnhorst\\
 UC Santa Cruz, 1156 High St, Santa Cruz, CA, 95060}
 

\maketitle


\begin{abstract}
Computer science theory provides many different measures of complexity of a system including Kolmogorov complexity, logical depth, computational depth, and Levin complexity. However, these measures are all defined only for deterministic Turing machines, i.e., deterministic dynamics of the underlying generative process whose output we are interested in. Therefore, by construction they cannot capture complexity of the output of stochastic processes - like those in the real world. Motivated by this observation, we combine probabilistic Turing machines with a prior over the inputs to the Turing machine to define a complete stochastic process of Turing machines. We call this a stochastic process Turing machine. 
We use stochastic process Turing machines to define a set of new generative complexity measures based on Turing machines, which we call stochastic
depth. As we discuss, stochastic depth is related to other such measures including Kolmogorov complexity and Levin complexity.  However, as we elaborate, it has many desirable properties that those others measures lack. In addition, stochastic depth is closely related to various thermodynamic properties of computational systems. Stochastic process Turing machines and stochastic depth allow us to study complex, stochastic systems like the human brain, societies, and evolution all from within the framework of formal computation.
\end{abstract}

\section{Introduction}\label{Motivation}
Kolmogorov complexity \cite{solo64,kolm65,livi08} has been a central concept in computer science theory since its inception, and it arguably created algorithmic information theory (AIT). It has spawned a zoo of complexity measures like logical depth \cite{bennett1995logical}, Levin complexity \cite{LEVIN198415}, resource-bounded Kolmogorov complexity \cite{ALLENDER201114}, and computational depth \cite{antunes2006computational}. (See \cite{feldman1998survey} for a survey of many measures.) It has found wide use in science, ranging from computer science \cite{livi08} and communication \cite{Kaplan2009}, all the way to biology \cite{Johnston2022} and physics \cite{BAEZ2012, tadaki2008statistical}.
The prefix-free Kolmogorov complexity (usually referred to as just Kolmogorov complexity) $K(x)$ of a string $x$ is defined as the length of the shortest program $p$ running on a universal 
Turing machine (UTM) that outputs $x$. This is the formal definition for the idea that, given a computational process, the complexity of an object is how much information is needed as input to produce the object.


Just as there are many complexity measures, there are many generalizations of TMs, like prefix TMs, UTMs, nondeterministic TMs\footnote{Not to be confused with PTMs}, and probabilistic TMs (PTM) (see the appendix of \cite{wolpert_2024_self-sim} for a review of TM theory). In particular, a PTM is a TM with multiple update functions, which randomly chooses one at each iteration, usually uniformly. PTMs are extremely powerful for modeling real phenomena since we live in a (effectively) probabilistic world.

{However, $K$ has an intrinsic downside -- it doesn't depend on the length of the computational path that produces the output from the minimal length input. Resource-bounded Kolmogorov complexity was designed to fix this \cite{LEVIN198415,ALLENDER201114,lu2022optimalcodingtheoremstimebounded}. A classic example is Levin complexity, which adds a term that penalizes the runtime of the algorithm:
$$
Kt(x)=\min_{p}\{|p|+\log |t|\}.
$$
However, there remains a subtle yet significant fault with $Kt$ and other resource-bounded complexities (not to mention logical depth and $K$), namely that they do not take into account how invertible or non-invertible the steps of the computation are.

To convey why this is problematic, let $T$ be a deterministic TM. A step between states $a$ and $b$ is 1-to-1 if there is only a single configuration of $T$, namely $a$, that leads to $b$. A step is many-to-1 if there are many such $a$ that produce $b$.  We note that, in many senses, if you have a step during the trajectory of $T$ that's invertible, \textit{it isn't actually a cost}. This is explicitly the case in thermodynamics, when one considers a physical implementation of the TM. If a step is a 1-to-1 (invertible) rather than a many-to-1 (non-invertible), there is much less thermodynamic cost \cite{wolpert2019stochastic}. Surprisingly, this fault with resource-bounded complexity measures can be solved with Bayesian reasoning.

To formally address this issue, we start by introducing some notation and Bayesian concepts. TMs provide a natural forward conditional probability, which is the probability that a TM outputs $y$ when given an input $x$. The forward conditional is written as $\pi(out=y\,|\,in=x)$, and for a deterministic TM this is
a Kronecker delta, $\pi(out=y\,|\,in=x)=\delta(U(x),y)$. To compute the inverse conditional with Bayes' theorem, a prior is needed. AIT provides a canonical prior: the fair-coin measure,
$\pi(in=x)\propto 2^{-|x|}$. However, we can generalize this to an arbitrary prior. A PTM equipped with such a prior is bona fide stochastic process, which we call a stochastic process Turing machine (SPTM).

Given this notation, we can express the information theoretic way to quantify how non-invertible a step of the TM is. If the step is from $t$ to  $t+1$, this quantity is the
backward conditional entropy, $S(X_t \,|\, X_{t+1})$ (where $X$ is the random variable of the full instantaneous
description of the TM, i.e. its state). If this conditional entropy is maximal, then knowing the state of the 
TM at $t+1$ tells you \textit{nothing whatsoever} concerning the preceding state --- which in turn means that
the (deterministic) map from that preceding state $x_t$ to the state $x_{t+1}$ must be maximally many-to-1.

In addition, the longer the trajectories are --- exactly the possibility that time-bounded versions of Kolmogorov complexity is concerned with ---
the larger the sum over the trajectories of the conditional entropies $S(X_t \,|\, X_{t+1})$. In other words, that sum
captures \textit{both} how many-to-1 the dynamics of the TM is, \textit{and} how long the trajectories are that the TM produces.

After accounting for the fact that a trajectory of length $k$ generated by running a TM is a random variable, 
these considerations motivate the following measure:

\eq{-\sum_{k = 1}^\infty \pi(k) \sum_{\tau : |\tau| = k} \sum_{i=2}^{k} S(X_{i-1} \,|\, X_{i}),
\label{eq:1X}
}
where for all $i$, $X_i$ is the random variable over possible states of the TM whose marginal distribution is given by $\pi_i$. 
So \cref{eq:1X} is the expectation value (over trajectory lengths) of the sum along a trajectory of the backward conditional entropy of each step in that trajectory. However, this is very difficult to compute (let alone in the computer science sense of computability), so we consider the case in which the steps of the TM are likely to be close
to invertible. Then one would expect that we can expand \cref{eq:1X} as
\eq{
-\sum_{k = 1}^\infty \pi(k) \sum_{\tau : |\tau| = k} \sum_{i=2}^{k} S(X_{i-1} \,|\, X_{i}) \simeq -\sum_{k = 1}^\infty \pi(k) S(X_1 \,|\, X_k, k).
\label{eq:2X}
}
In the limit that every step is \textit{exactly} invertible, this approximation becomes exact since
all conditional entropies go to zero. In turn, with abuse of notation, we can define 
$X_{halt}$ to be the random variable over the states that the TM can have when it halts, and we can then
rewrite the RHS of \cref{eq:2X} as 
\eq{
\label{eq:3X}
&S(X_1 \,|\, X_{halt}) = -\sum_{x_1, x_{halt}} \pi(in = x_1, out = x_{halt}) \log \pi(in = x_1 \,|\, out = x_{halt}) \\
&= -\sum_{x_{halt}} \pi(out = x_{halt}) \left[
\sum_{x_1} \pi(in = x_1 \,|\, out = x_{halt}) \log \pi(in = x_1 \,|\, out = x_{halt}) \right].
\label{eq:4X}
}
In other words, the sum of the conditional entropies of each step going backwards on a trajectory is approximately equal to the conditional entropy of the beginning state of the trajectory, given its ending state.

\cref{eq:3X} implies that measures like those in \cref{eq:2X}, namely those that take into account how non-invertible steps in a trajectory are, might be naturally expressed in terms of quantities like $- \log \pi(in = x_1 \,|\, out = x_{halt})$.

This motivates our definitions of stochastic depth, and in fact the RHS of \cref{eq:3X} is just the expected value of what we define as the ``stochastic depth of generating $x_{halt}$ from $x_1$." This specific version of stochastic depth is exactly equal to the backward conditional $-\log \pi(in = x \,|\, out = y)$, and we define related stochastic depth measures that vary this by considering sums over inputs, maxima over inputs, and the mode of the output distribution.

The inner sum in \cref{eq:4X} is upper-bounded by
\eq{
\max_{x_1} \left(-\log \pi(in = x_1 \,|\, out = x_{halt})\right),
}
and that bound is tight for all $x_{halt}$ such that $\pi(in = x_1 \,|\, out = x_{halt})$ is peaked.
This upper bound is precisely an important variant, which we call $S_2$, of the above stochastic depth.


We can motivate stochastic depth even further by describing an important algebraic property that many of its variants satisfy. A measure often considered in AIT that has similarities to the conditional entropy is the conditional Kolmogorov complexity, written $K(x\,|\,y)$. Infamously, conditional (deterministic) Kolmogorov complexity does not obey the simple properties of conditional Shannon entropy, namely that $H(Y\,|\,X)=H(X,Y)-H(X)$. It is well known that $K$ satisfies these analogously only up to a logarithmic term \cite{livi08}:
\eq{|K(x,y)-K(x)-K(y\,|\,x)|=\Omega(\log K(x)).}
In contrast, the ``stochastic depth of generating $x_{halt}$ from $x_1$'' avoids this issue by construction.

In section 2, we formally define SPTMs, the output of an SPTM, and the prefix version of PTMs. 
In section 3, we define a few variants of stochastic depth and discuss their computability and relations to thermodynamics. In section 4, we define a complexity class for SPTMs and discuss the relation to one-way functions. In section 5, we discuss the applications of SPTMs and stochastic depth. These include a recent, fiery literature debate on assembly theory \cite{{kempes2024assembly,Sharma2023,Uthamacumaran2024, abrahao2024assemblytheoryapproximationalgorithmic, ozelim2024assemblytheoryreducedshannon}} and a formalization of the ``stepping stone" phenomenon in evolution.

\section{SPTMs}\label{SPTMs}
In this paper, we adopt all the notational conventions of elementary algorithmic
information theory. All variables are implicitly encoded in the conventional way as bit strings~\cite{livi08} and $\langle z \rangle$ is used
to make the encoding of a positive integer $z \in \N$ explicit. For any
bit string $y$, $|y|$ denotes its length. Similarly, we write $\langle x, y \rangle$ for the conventional
encoding of two (positive integer) numbers as a single bit string~\cite{livi08}. For any string $\vec{z}$,
$z_j$ is its $j$'th element, and $\vec{z}_j$ is the substring from index $j$ to the end of $z$.
For simplicity, we consider stochastic versions of prefix TMs, which are implemented in the
usual way using a read-only irreversible input tape, a write-only irreversible output tape, and an arbitrary
number of intermediate work tapes with no such restrictions. All tapes have only binary symbols (no blank symbol), and any work tapes are initially filled with zeros, as is the output tape. 
We say that such a machine $T$ halts on input $x$ with output $y$, and write $T(x)=y$, if when $T$ halts
$x$ is to the left of the input head and $y$ is to the left of the output head. 
Essentially by definition, the set of $x$ on which $T$ halts forms a prefix code. The set of input strings to such a TM is written as $IN$ with elements $in$, and the set of output strings is written as 
$OUTPUT$, with elements $output$.
The state of the output string at iteration $t \in \N$ is written as $output_t$. (The input string
never changes from one iteration to the next.)

We define a \textbf{prefix} probabilistic Turing machine (PPTM) to be a prefix TM that flips a uniform coin at 
each iteration to choose between two pre-specified update functions. It is a conventional PTM that
happens to be a deterministic prefix TM if its two update functions are identical. In the case of a PPTM, we overload notation and use $IN$ to indicate the random variable of the 
input strings (assuming it is generated stochastically), and $OUTPUT$ to indicate the random variable of the
output strings.

Note though that even though a PPTM uses the architecture of a prefix Turing machine, the set $H_U$ of all
inputs $x$ such that there is nonzero probability that the PPTM $U$ halts with its input head 
immediately to the right of $x$ will \textit{not} form a prefix set in general. This is true even though
it is defined using an ``architecture'' (the set of tapes and how they co-evolve) of a prefix TM.
The reason for this is that
there might be some probability $q > 0$ that the PPTM halts with $x$ immediately to the left
of the input head, but also a probability $r > 0$ (where $1-q \ge r $) such that the PPTM continues
reading symbols after reading input string $x$, and then halts with some string $xs$ to the left of its head,
where $s$ is not the empty string. In the following, for any PPTM $U$, we write
$\U$ for the countably infinite set of possible IDs of $U$.
We adopt the convention that the initial iteration of the PPTM is $t = 0$.

An $N$-\textbf{function} PPTM is an extension of a PPTM to endow it with $N \ge 2$ update
functions. The PPTM is run by uniformly randomly choosing among those $N$ update functions at each iteration.
We can use a conventional $2$-function PPTM to run an algorithm that only flips a coin at certain points in its
program.\footnote{To implement this with a PPTM, construct the read head to have a special ``flip'' state, 
and have that differ from its initial state. Then have that head jump into the flip state if and only if
has just a read a special sequence of bits on the input tape (which correspond to the instruction in the program, ``now
flip a bit''). Finally, have the two update functions be identical if that head is in any state \textit{other} than the flip state.} 
This means that for any fixed $n$, we can use a $2$-function PPTM to implement
a PPTM where the probability distribution of the coin has any desired bias given by a real number with up to $n$
bits. (See Lemma 7.12 of \cite{arora2009computational}.) However, the only advantage of considering PPTMs with $N > 2$ update functions would be
to improve the time-bounds of the PPTM, i.e., for computational complexity reasons. Accordingly, except where
explicitly noted otherwise, we take $N=2$ . (The definitions can be extended to $N > 2$ in a straightforward
way.)

A \textbf{universal} $2$-function PPTM $U$ is one that takes an argument
$\langle \delta^1, \delta^2, x, t \rangle$
where $\delta^1, \delta^2$ are the updates functions of any two universal prefix TMs, $x$ is an input string,
$\langle \ldots \rangle$
is any convenient coding of those three quantities into a single bit string,
and $t$ is the iteration number of the PPTM the UPPTM is emulating. $U$'s two update functions are just
the two rules, ``implement the first function in the input $\langle \delta^1, \delta^2, x \rangle$'', or ``implement the
second function in the input $\langle \delta^1, \delta^2, x \rangle$'', respectively, with some overhead
for decoding $U$'s second argument, $\langle \delta^1, \delta^2, x \rangle$, choosing between the two update functions specified
in that argument, etc.
By taking $\delta^1 = \delta^2$, we can have $U$ emulate any deterministic TM. 
The halting problem for any such SPTM $U$ is undecidable by that machine. 



Note that the stochastic update rule of a PPTM is encoded by $2$ update rules,
and each of those update rules is a finite string. Accordingly, the argument list (i.e., input string) of any universal PPTM is
a finite bit string. So the set of all such input strings to a universal PPTM is countable. In contrast, there
are an uncountable number of (real-valued) conditional distributions over $\U$.
Accordingly, we could not encode an arbitrary such conditional distribution as a finite bit string that
could be input to a universal computational system of any sort. 


A \textbf{full} PPTM is one such that for any given bit string
$x$, there is at least one input to the PPTM such that it halts for that
input with nonzero probability of having $x$ on its output tape. This contrasts with the case considered in the 
standard computational complexity definition of a PTM, in which with probability $1$
the output tape contains either the single-symbol string `1' or the single-symbol string `0' when the
PTM halts. Note that for a PPTM to be full implicitly
means that for any counting number $n$, there is nonzero probability of the PPTM running for at least $n$ iterations
for some associated input.

As shorthand write the (time-invariant) update conditional distribution of PPTM $U$ as 
\eq{
\pi_U(u_{t+1} \,|\, u_t) 
}
or just $\pi(u_{t+1} \,|\, u_t)$ when $U$ is clear from context.
For completeness, we specify that for any
ID $u_t  \in \U$ in which the PPTM is in its halt state, and any $u_{t+1} \in \U$,  $\pi_U(u_{t+1} | u_t) = \delta(u_{t+1}, u_t)$,
where $\delta(\cdot\,,\cdot)$ is the Kronecker delta function. This is the (stochastic) update rule for the PPTM $U$.
$U$ defines a (discrete time) partial Markov chain over $\U$. 

Define some \textbf{prior} semi-distribution $\pi(u_0)$ over the initial ID of the PPTM, $u_0$.
Combined with $\pi(u_{t+1} \,|\, u_t)$, such a prior fully specifies a stochastic process, $\pi(u_{t_1}, \ldots)$, defined
for any finite list  $\{u_{t_1}\}$ of (at least one) iteration-indexed IDs of the PPTM.
Such a stochastic process is called an SPTM.  As an example, if $\pi(u_0)$ is a ``discrete Lebesgue measure''~\cite{livi08} defined over
a prefix-free set of bit strings that encodes all integers, then it is normalized. 
We don't assume such normalization in general though. Note that
if the semi-distribution $\pi(u_0)$ is not normalized, neither is the marginal distribution at later steps.

As a point of notation, we write $in$ for the (iteration-independent) string on the input tape
of the PPTM $U$, formed by sampling $\pi(u_0)$.
In addition, with abuse of notation, an SPTM $(\pi(u_0), U)$ for a PPTM $U$ and prior $\pi(u_0)$ will 
often be written as just $\pi$.
Abusing notation further, $\pi(u_t)$ will mean that SPTM evaluated at iteration $t$, and $\pi(in)$ will  mean the prior
over initial inputs. (Note that the prior $\pi(u_0)$ over initial IDs is just $\pi(in)$ times associated delta functions,
setting the initial strings of the tapes other than the input tape, the initial state of the head and the associated pointer(s).)

Write $\tau$ for the random variable of the first iteration in a particular run of the SPTM at which $U$ is in its halt state,
and $\pi(\tau = i)$ for the probability that that smallest iteration equals $i$. In general,  there may be
some set $A$ of input strings that has nonzero probability under the prior such that
$U$ never halts so long as the input is from $A$. As a result, in general
\eq{
\sum_{i \in \N} \pi(\tau = i) &= \sum_{i \in \N, in} \pi(\tau = i \,|\, in) \pi(in) \\
 &\le 1.
}
So $\pi(\tau)$ is a semi-distribution over $\N$, in general. Write $output_t$ for the random variable of the contents of the output tape of the SPTM
at iteration $t$. 
So  
\eq{
out := output_\tau
} is the random variable giving the string on the SPTM's output tape if and
when it reaches the halt state. In more abuse of notation, we write $\pi(out = i)$
for the probability that $out = i$, and $\pi(out = i \,|\, in = j)$ for the conditional probability
that the SPTM ends with its output set to $i$, given that it started with its input set to $j$. Note that since we are considering stochastic \textit{prefix} probabilistic TMs, the input variable
and the output variable both exist when the computation halts, just on different tapes.

The \textbf{output} of an SPTM $(\pi(u_0), U)$ on input $x$ is independent of the prior, given by 
\eq{
	{\hatout}(x) &:= {\rm{argmax}}_y Pr_U(out = y \,|\, in = x) \\
		&= {\rm{argmax}}_y \sum_{t \in \N} Pr_U(output_t = y \,|\, in = x, \tau = t) Pr_U(\tau = t \,|\, in = x)
\label{eq:6}
}
assuming that with nonzero probability, the PPTM halts when run on input $x$, i.e.,
that there is nonzero probability under the SPTM of the event $(in = x, \tau=t)$
for at least one $t \in \N$. (Throughout this paper we abuse notation and write expressions like
``$\pi(in)$'' as shorthand for $\pi_{IN}(in)$, the distribution over inputs evaluated for the input $in$.)

We take the output to be undefined for some input $x$
if with probability $1$ the PPTM $U$ never halts for input $x$, or if it halts with nonzero probability, but the mode of
$Pr_U(out \,|\, in = x)$ is not unique. Accordingly, we write $Pr_U(\hatout(x) = y)$ for the probability that 
$U$ eventually halts on input $x$, 
and conditioned on its halting, $Pr_U(out \,|\, in = x)$ has a unique maximum, at $out = y$. 

Note the consistency of this definition with the standard computational complexity definition of the output of a 
conventional PTM, which
applies to a special case where the PTM always halts and is not full, with probability $1$ outputting either a $1$ or a $0$. 
Of course, like with PTMs, we could instead define the output of a PPTM to always exist, no matter what the distribution
over outputs conditioned on an input, by adopting some convenient tie-breaking mechanism.

We could make the connection with the definition of conventional PTMs even tighter.
For example as in the definition of the class $\rm{\textbf{BPP}}$, we could require that for any input $x$ the
PPTM reach a halt state with probability $1$ within ${\cal{O}}(P(|x|))$ iterations, where $P(|x|)$ is a polynomial 
function of $|x|$. Also as in the
definition of  $\rm{\textbf{BPP}}$, to ensure ``robustness'' in the output that is generated, we could modify \cref{eq:6} to have the output
be undefined if  $ {\rm{argmax}}_y Pr_U(out = y \,|\, in = x)$ is not at least $1/|x|$ greater than the output
with the next largest output probability.

As discussed above, $H_U$, the set of all inputs $in$ such that there is nonzero probability
that $U$ halts on input $in$, is not a prefix-free set in general, even though the underlying TM architecture of $U$ is
that of a prefix TM. However, note that the set $H_U$ is a superset of a union of prefix-free sets, namely
the prefix sets of inputs of each of the 
update functions defining $U$. Accordingly, we can choose any one of those 
prefix-free halting sets of one of the update functions, 
$H_{\delta_i} \subseteq H_U$, and then define $IN_U$ to be the set of all elements of $H_U$.

Suppose we have two UTMs, $U$ and $U'$. Suppose we also have a prior distribution $\pi$ with a promise that its
support is restricted to strings $x$ such that both $U$ and $U'$ halt. So, the support of $\pi$ is a subset of the intersection
of the halting sets of $U$ and $U'$, $H_U \cap H_{U'}$. The field of average-case complexity has many very powerful results about the time and space complexity of running any
single such UTM over such a $\pi$. These results do not change if we choose $U$ or $U$', since the universality
theorem of UTMs means that the associated time-complexities differ only by an additive constant. 

However, suppose we pick one of those two UTMs, say  $U$. Then if we
change between two distributions $\pi_1$ and $\pi_2$ for that fixed UTM, even if both of those distributions have support restricted to
$H_U \cap H_{U'}$, then the resultant distribution over outputs of $U$ can change radically. Moreover, exactly \textit{how} that distribution
changes as we move between $\pi_1$ and $\pi_2$ can itself vary greatly depending on the relationship between $U$ and $U'$.

There is far more richness when the input string is sampled from a random distribution than has been investigated in the field of average-case complexity.

Similarly, suppose we fix the input distribution $\pi$, and consider a single SPTM, with two associated update functions
$\delta_1$ and $\delta_2 \ne \delta_1$. The time-complexity of moving between the two
deterministic TMs for update functions $\delta_1$ and $\delta_2$ changes by just an additive translation constant.
If we change the second update function $\delta_2$, then all we do is change the value of that additive constant.
On the other hand, changing the second update function $\delta_2$ can drastically affect the output distribution of the 
associated SPTM. 
In general, translating between TMs is far more consequential when considering SPTMs than it is when considering conventional deterministic TMs.

From now on we assume that each of the update functions, by itself, gives a (necessarily deterministic) UTM.
Therefore for both update functions $\delta_i$,
for  all finite output strings $y$, there is an element of $H_{\delta_i}$ that results in the associated deterministic
UTM producing $y$ and then halting. We also assume 
that the support of the prior $\pi(IN)$ is precisely the associated set $H_{\delta_i}$ for one of the
two update functions $\delta_i$, and so from
now on that it is what is meant by ``$IN_U$''. Note that even though there is nonzero
probability that SPTM $U$ halts on each $in \in IN_U$, in general there will also be nonzero probability that $U$
\textit{doesn't} halt on that $in$. So restricting to input strings in $IN_U$ doesn't mean we're forcing $U$
to always halt.

Unless explicitly stated otherwise, we assume that the relative
probabilities of the strings in $IN_U$ are given by the Cantor measure used to
define the ``universal a priori probability'' in algorithmic information theory~\cite{livi08}.
So for any two $s, s' \in IN_U$, the ratio of the prior values is
\eq{
\dfrac{\pi(u_0 = s)}{\pi(u_0 = s')} &= 2^{|s'| - |s|},
\label{eq:7}
}
where $|x|$ is the length of any bit string $x$.

As required, since $IN_U$ is a prefix set, by Kraft's inequality 
\eq{
\sum_{in} \pi(in) := \sum_{in \in IN_U} \pi(in) \le 1.
}

There are many different systems that are computationally universal. One can ``translate''
between any two of them at a cost of an overall, additive translation constant. (As an example, that is
the length of a compiler program that converts programs written in one computer language to
equivalent programs written in a different language.) 

It is commonly argued that this means that
we can meaningfully talk about ``the'' Kolmogorov complexity of a string, by simply ignoring the additive
translation constant that relates the length of an input $p_1$ for some universal Turing machine $T_1$ to the associated
input $p_2$ that would cause a different universal Turing machine, $T_2$, to compute the same output as
the one that $T_1$ produces when run on $p_1$.

However, even if one accepts that this additive constant has limited significance in algorithmic information complexity 
(the study of topics like Kolmogorov complexity), it has massive implications when investigating SPTMs. To see
this, first note that a semi-distribution like the one defined in \cref{eq:7} will change if we change the underlying probabilistic
UTMs. Moreover, we can measure that change with quantities like the  Kullback-Leibler divergence 
between the two distributions. (Or more precisely, bounds on that divergence, since in general it is uncomputable.)

Since we are considering full SPTMs, for any fixed $y$ the
Bayesian inverse
\eq{
\pi(in = x \, |\, out = y) &= \dfrac{\pi(out = y \,|\,in = x)\pi(x)}{\sum_x \pi(out = y\,|\,in = x)\pi(x)}
\label{eq:12}
}  
is well-defined and properly normalized to $1$, even if the prior $\pi(u_0)$ is
not. Similarly, $\pi(in = x \, |\, \widehat{out }= y)$ is properly
normalized, so long as $x$ is a value (of which there might many) such that $\hatout(x) = y$.


We can define
\eq{
\pi(\hatout = y) &:= \sum_{in : \hatout(in) = y} \pi(in)
\label{eq:1}
}
so that
\eq{
\sum_y \pi(\hatout = y) &= \sum_y \sum_{in : \hatout(in) = y} \pi(in) \\
	&\le\sum_{in} \pi(in) \le 1,
}
where the first inequality uses the fact that in general there is nonzero probability that the PPTM will never
halt if it starts with input $in$, and the second inequality uses Kraft's inequality. Similarly,
\eq{
\sum_y \pi({out} = y) &= \sum_y \sum_{in} \pi(out = y \;|\; in)\pi(in) \\
	&\le \sum_{in} \pi(in) \le 1.
}

We define an SPTM $U$ to be \textbf{total} if, for all inputs $in \in IN_U$, $\sum_y \pi(out = y \,|\, in) = 1$. An SPTM can be total
even if neither of the two deterministic TMs defined by its two update functions are total. To see how,
write any sequence of specifications of one or the other of its two update functions as $\vec{\delta} = (\delta(1), \delta(2), \ldots)$.
So every $\vec{\delta}$ is a bit string,
and the sum-probability of sequences of length $m$ which have not yet halted is lower-bounded by $2 \times 2^{-m}$.~\footnote{To
see this, just note that the probability that every element in that sequence is generated using the exact
same one of the two update functions of the TMs is $2^{-m}$, and by hypothesis, the associated TM cannot halt.}
Then the SPTM will be total if this lower bound is a strict equality in the limit of infinite $m$.

We will {not} assume that the SPTM is total unless explicitly stated otherwise. In particular,
an SPTM need not be total even though the support of the prior is restricted to the prefix-free set $IN_U$, a set of inputs for which 
there is nonzero probability that the SPTM halts.

\section{Stochastic Depth}\label{Stochastic Depth}

The key is that SPTMs allow us to define the probability of an input 
to a computation conditioned on the output of that computation via the Bayesian inverse
of the output conditioned on the input. We now formally notation to the quantity emphasized in Sec. \ref{Motivation}. The \textbf{stochastic depth of generating $y$ from $x$} is
\eq{
s_\pi(x\rightarrow y) := -\log \pi(in = x \,|\, out = y).
}
Note that we do not impose the requirement that $y$ be the mode of the output for input $in = x$, for the
simple reason that it might not be. Also note that $s_\pi(y \rightarrow x)$ does {not}
involve the Bayesian inverse of $s_{\pi}(x \rightarrow y)$.

As for variants of the above, there are (at least) three ways to use the backward conditional
to quantify  the ``difficulty'' or ``complexity'' of a particular output string. Though ultimately we will settle on a single definition, for now we will refer to all of these slightly distinct
measures as the \textbf{stochastic depth} of a string $y$ under some SPTM $\pi$:
\eq{
S_0(y) &:= -\log \left(\max_{x : {\hatout}(x) = y} \left[\pi \left(in = x\right)\right] \right) \\
S_1(y) &:= -\log \left(\max_x \left[\pi \left(in = x \,|\, {\hatout}(x) = y\right)\right] \right) \\
S_2(y) &:= -\log \left(\max_x \left[\pi \left(in = x \,|\,  out = y \right)\right] \right),
}
where all logarithms are base $2$, as conventional in computer science and information theory.
(These definitions of ``stochastic depth'' should not be confused with the concept of that name that
is used in machine learning.)

The difference between the last two alternatives is that $S_2(y)$ does not suppose that
$y$ is the mode
of the conditional distribution of $out$ given $in = x$, unlike $S_1(y)$. 
\textit{A priori}, $S_1(y)$ could be larger than $S_2(y)$ or vice-versa,
depending on $y$ and $U$. Note that by the assumptions concerning the support of $\pi(IN)$, $S_0$ must be 
well-defined for any finite string $y$. It is straightforward (though cumbersome) to impose additional assumptions concerning 
the update functions to ensure that $S_1$ and $S_2$ are always well-defined as well.

There are many complexity measures involving SPTMs that
are related to $S_0, S_1$ and $S_2$ that might also be of interest. Most obviously,
we could replace the max's in those definitions with sums:
\eq{
\overline{S}_0(y) &:= -\log \left(\sum_{x : {\hatout}(x) = y} \left[\pi \left(in = x\right)\right] \right) \\
\overline{S}_1(y) &:= -\log \left(\sum_x \left[\pi \left(in = x \,|\, {\hatout}(x) = y\right)\right] \right) \\
\overline{S}_2(y) &:= -\log \left(\sum_x \left[\pi \left(in = x \,|\,  out = y \right)\right] \right).
}
For all $y$, for all $x$ such that $\hatout(x) = y$,
\eq{
\pi(in = x \,|\, \hatout(x) = y) \ge \pi(in = x).
}
So for all $y$,
\eq{
\overline{S}_1(y) \le \overline{S}_0(y)
}
and
\eq{
{S}_1(y) \le {S}_0(y).
}


There are several different ways to motivate these measures as  interesting
characterizations of a string $y$ (given some fixed, universal SPTM). Perhaps most obviously,
given our choice of the fair-coin measure as the prior over inputs, $S_0(y)$ is the
minimal size (measured in bits) of any input $x$ that \textit{might} have generated $y$. So it
reduces to Kolmogorov complexity if we have a deterministic universal SPTM, i.e., a universal SPTM whose
update functions are the same. Similarly, for a deterministic universal SPTM, $2^{-\overline{S}_0(y)}$ is the universal a priori probability of $y$, i.e., $\pi(out = y)$~\cite{livi08}.

$S_1(y)$ and $S_2(y)$ refine this complexity measure. To see this, first re-express $S_0(y)$ as
the (negative logarithm of) the prior probability of any $in$ that (with greatest probability) 
would have produced $y$ (up to
$y$-independent additive constants). An obvious refinement would be to replace the prior
probability in this definition of $S_0(y)$ with the \textit{posterior} probability, conditioned on $y$.
With this modification, we have a measure of  the computational
difficulty of ``going backwards,'' undoing the SPTM, to generate $x$ from the output $y$.
$S_1(y)$ and $S_2(y)$ are two versions of this refinement of Kolmogorov complexity.
From now on, the phrase ``stochastic depth'' will only be used for measures $S_1(y)$ and / or $S_2(y)$.

Further insight into the nature of stochastic depth comes from noting that in the case of a deterministic SPTM, $S_1(y)$
and $S_2(y)$ become identical functions, reducing to
\eq{
-\log \left[\frac{2^{-K(y)}}{\sum_{x : U(x) = y} 2^{-|x|}}\right] &= K(y) + \log  \left[\sum_{x : U(x) = y} 2^{-|x|}\right],
\label{eq:4}
}
where $K(y)$ is the Kolmogorov complexity of $y$ under the (deterministic) TM $U$
and the condition $U(x) = y$ means that $U$ produces $y$ and then halts when initialized
with $x$ on its input tape. 

The first term on the RHS of \cref{eq:4} tells us that stochastic depth increases with the computational difficulty of producing $y$ (as quantified by
Kolmogorov complexity) from that single input $x$ that has the least such difficulty. However, that term, by
itself, doesn't account for how \textit{many} inputs produce $y$. It doesn't account for the possibility that
while no single simple input produces $y$, there are a vast number of inputs, none simple by itself,
that produces $y$. 

Stochastic depth corrects this shortcoming, by adding a term to Kolmogorov
complexity that shrinks the more inputs there are that produce $y$ (where those inputs are
weighted by their associated prior probabilities). This second term is just (log of) the universal a priori probability of $y$,
so by Levin's coding theorem, these two terms on the RHS of \cref{eq:4} are identical up to $O(1)$, i.e., up to a function
of $y$ that does not grow larger than some finite upper bound as one varies $y$.

In~\cite{livi08}, for any universal deterministic TM $U$, the \textbf{universal probability} of $y$ is defined as
\eq{
Q(y) &:= \sum_{x : U(x) = y} 2^{-|x|} \\
	&= \sum_{x : U(x) = y} \pi(x) \; = \; 2^{-\log \overline{S}_0(y)}
}
where $Q(y)$ is referred to as the ``algorithmic entropy'' of $y$ \cite{baez2012algorithmic}.
We can now note that for a deterministic SPTM $U$, for all $x, y$, such that $U(x) = y$,
\eq{
\pi(in = x) \;\le\; \dfrac{\pi(in = x)}{Q(y)} \;=\; 2^{-s_\pi(x\rightarrow y)} 
\;\le\; 1,
}
where the equality uses \cref{eq:12} and the fact that for a deterministic SPTM, $\pi(out = y \,|\, in = x) = \delta(y, U(x))$. 
Applying Levin's coding theorem to \cref{eq:4} tells us that
in the deterministic limit, up to $y$-independent constants, the stochastic depth of a string $y$  is proportional to $-\log m(y)$, negative
log of the \textbf{universal measure} of $y$, plus the log of the universal probability of $y$~\cite{livi08}.

This result means that in the
deterministic limit, up to an overall additive constant (namely, log of Chaitin's $\Omega$),
the stochastic depth of a string $y$ is precisely the minimal heat that must be expelled into the
environment in order to generate the string $y$ on any physical system that evaluates the 
partial function of the Turing machine, assuming there is no change in the expected energy of that physical
system as it carries out that evaluation.\footnote{To 
be precise, it is the sum of the Landauer cost of evaluating the partial function defined by the TM,
plus the mismatch cost of implementing that partial function, where  for calculating the mismatch cost,
i) we set the ``prior'' (in the language of stochastic thermodynamics) 
to the fair-coin measure; ii) the actual initial distribution over the states of the
TM is a delta function about the input $x$ with maximal probability under the prior out of all of
those inputs that map to the desired output $y$.  See Sec. XIV of~\cite{wolpert2019stochastic}.} 
So in the case of a single TM that is driven by coupling
with a thermal bath at temperature $T$, the stochastic depth is proportional to the minimal heat that
is produced by generating $y$ on the given TM. (The ``minimum'' in this case being a minimum over all
input strings $x$.)

It was informally argued in~\cite{zure89a} that the minimal ``thermodynamic cost''
to transform a bit string $x$ into a bit string $y$ using a TM $U$ is lower-bounded by 
\eq{
K(x \,|\, y) := \min_{p : U(\langle p,y\rangle) = x} |p|,
\label{eq:5}
}
the conditional Kolmogorov
complexity of $x$ given $y$ as side-information~\cite{livi08}. This argument was formalized, corrected, and
partially validated in Sec. XIV of~\cite{wolpert2019stochastic}, by identifying the ``thermodynamic
cost'' with the minimal heat expelled to the environment in a process that sends $x$ to $y$. (See also~\cite{kolchinsky2023generalized}.)

We can apply that result here, to lower-bound the ``thermodynamic cost'' of generating $y$ from 
any input string $x$. Suppose we evaluate that lower bound, $K(x\,|\,y)$,
where $x$ is the minimal length input string that generates $y$.
Since $x$ uniquely specifies $y$ (just by running the TM starting from $x$), $K(\langle x \rightarrow y\rangle) = K(x)$
up to an additive constant.
In turn, since $x$ is incompressible, $K(x) = |x|$ (up to additive constants). But $|x|$ is just $K(y)$.
So the expression in \cref{eq:5} is zero, up to an additive constant. In fact, evaluating that constant might
give a negative value, depending on the details of the deterministic SPTM being used to generate $y$ from $x$.
(For example, that would be the case if extra instructions need to be included in the original program $x$
to have erase itself after generating $y$.) In contrast, both $S_1(y)$ and $S_2(y)$ will
be strictly positive in general --- no matter what the underlying deterministic SPTM is. So even though
$K(x \,|\, y)$ might be negative, the actual minimal entropy flow is strictly positive, being lower-bounded
by the strictly positive stochastic depth.
This illustrates how weak the lower bound in~\cite{zure89a} can be.\footnote{On the other hand, it is important to
note that while stochastic depth is strictly positive, the expression in \cref{eq:5} has a finite upper bound as
one varies $y$, i.e., it is bounded by a positive constant, and in this sense is similar to the bound
of $K(x \,|\, y)$. See~\cite{wolpert2019stochastic}.}


\cref{eq:4} illustrates that, unlike logical depth~\cite{bennett1995logical},
computational depth~\cite{antunes2006computational} and Levin complexity~\cite{livi08}, stochastic depth says that it is not exactly the raw number of iterations to produce $y$ from $x$ that matters. 
Rather if stochastic depth is high and $K(y)$ is small, there are many (prior-weighted)
programs $x$ that all undergo many-to-1 maps to produce the same result, $y$. 

To illustrate this, suppose that like in the definition of one-way functions in computational
complexity theory, we use a prior that is uniform over all input of size (measured in bits) up to some finite $n$
(rather than the fair coin measure).
Again consider the limit that the SPTM is actually deterministic. In this case,
stochastic depth reduces to 
\eq{
\log \left(\sum_{x : U(x) = y} 1\right).
}
So the greater the total fan-in to all of the IDs on the trajectory from $x$ to $y$, 
the greater the stochastic depth. This agrees with the fact that large total fan-in
results in large stochastic depth for the prior based on the Cantor measure.

\section{Computational complexity and SPTMs}\label{ComputationalComplexityAndSPTMs}

One of the primary concerns of computational complexity theory is how
the resource costs to perform a computation scale with the size of the input to 
that computation~\cite{arora2009computational,sipser1996introduction}.
It is straight-forward to use our definitions to extend this concern to SPTMs and stochastic thermodynamics,
i.e., to the case where the resource cost is the minimal amount of heat that must be expelled to a heat
bath by any physical system that implements the SPTM, running on the given input to that SPTM.

We define that an SPTM $\pi$ is \textbf{energy-$T$} if
\eq{
2^{s_\pi(x\rightarrow\hatout(x))}= O\left(T(n(x))\right),
}
where $T(\cdot)$ is an increasing function from $\R^+ \rightarrow \R^+$,  $O(\cdot)$ is the usual big-oh operator, and we adopt the shorthand
\eq{
n(x) := -\log ({\pi(x)}) \,.
}
So in particular, $\pi$ is polynomial energy if
\eq{
2^{s_\pi(x\rightarrow\hatout(x))}= O\left(T(n(x))\right)
}
for a polynomial $T$. 


Another concern of computational complexity theory, closely associated with resource costs, is the (scaling of)
the amount of the resource needed by any TM to answer whether a given input string is
a member of a given language of input strings, where the TM is guaranteed to halt with its answer both
if the input string is in the language and if it is not.

Accordingly, here we define an SPTM $\pi$ to be a \textbf{decider} SPTM if for all $in$,
$\hatout(in)$ is either $0$ or $1$. (An alternative, stricter definition would require
that for all $in$, $\pi(out = y \,|\, in) = 0$ if $y \not \in \{0, 1\}$).

Paralleling the standard terminology for PTMs, 
we say that a language $L$ is in class \textit{SDIFF}$(T(n))$ iff there is a decider SPTM $\pi$ that has energy $\le T(.)$ for all input
strings, and if for all $in \in L$, $\hatout_\pi(in) = 1$.
The class \textbf{SP} is then defined as the language $\cup_{c \ge 1}$ \textit{SDIFF}$(n^c)$.


Define $B_\pi(n) := \{in : \pi(in) \ge 2^{-n}\}$. Also for any SPTM $\pi$, define
\eq{
\pi^{-1}(y) := \{in : \hatout(in) = y\}.
}
Note that in general, $\pi \left[in = \pi^{-1}(y)\right]$ can equal $0$, if there are no inputs that result
in a modal output of $y$. However, it is always the case that
\eq{
\pi\left(in = \pi^{-1}(\hatout_\pi(x))\right) \ge \pi(x)
}
since $p \in \pi^{-1}(\hatout_\pi(x))$.

These definitions allow us to define 
a \textbf{one-way SPTM} as a polynomial SPTM $\pi$
such that there is a negligible function $\epsilon(n)$ such that for all polynomial SPTMs $\pi'$, for all counting numbers $n$,
\eq{
\pi\left(in : \hatout_{\pi'}\left(\hatout_\pi(in)\right) \in \pi^{-1}(\hatout_\pi(in)) \,|\, in \in B_n\right) \le \epsilon_n.
}
One-way functions are conventionally defined using PTMs with a \textit{uniform} prior, while the prior is fully general here.

\section{Applications}\label{Applications}

A particularly salient application of this formalism is the recent literature debate in the study of complexity measures for the formation of life \cite{{kempes2024assembly,Sharma2023, Jaeger2024, cronin2024salientlimitationsonsalient, Uthamacumaran2024, abrahao2024assemblytheoryapproximationalgorithmic, ozelim2024assemblytheoryreducedshannon}}. Assembly theory (AT) is a recent framework for studying the complexity of physical and biological systems like molecules and evolutionary selection. The fundamental structures of AT are abstract \textit{building blocks}, which get recursively built to produce more complex objects. Once a more complex object has been assembled, it gets added back to the building block pool and is considered ``available for use" to construct further objects. The complexity is measured using the assembly $A$,  which computes how difficult it is to construct a given emergent object and depends on the number of copies of each building block and how difficult each building block was to build. AT starts to mimic biological selection by proposing a stochastic process in which each building block has a probability of getting chosen when constructing a new object. While the authors of AT argue that evolution can be seen as repeated, probabilistic runs of a TM, they do not formalize this notion or the ways in which AT describes a stochastic process.

It is argued that AT is a novel measure capable of providing a new bridge between biology and physics in the abstract, but it has been criticized on the basis that it falls under the umbrella of known LZ compression algorithms. Moreover, it is claimed that AT performs similarly or worse when compared to other common coding schemes.While these might be true, these criticisms seem to miss an important point: AT is designed to describe a stochastic process.
We do not take a side in the debate, but rather suggest that the SPTM framework is the proper setting to study AT. It is clear that AT is not yet formalized in the language of algorithmic information theory and much could be gained from doing so. This is necessary since criticisms of AT incorrectly apply complexity measures for deterministic computational systems to AT, which describes stochastic systems.

Relatedly, many have noted that it is often easier for natural or artificial selection to generate a complex system if it first evolves precursor, modular components which act as “intermediate stepping stones” for that evolution. For example, this is the basis for the semi-formal arguments motivating AT and for the semi-formal arguments made by Holland motivating the “royal road” hypothesis \cite{jones1994description}. Arguably, this line of reasoning relying on a “stepping stone effect” even extends back to Herb Simon \cite{simon1962architecture}.
We suggest that the stepping stone effect be formalized (arguably for the first time, in a broadly applicable way) using SPTMs. The goal would be to explain how a stochastic algorithm (e.g., natural selection) whose initial state was set by sampling a distribution $\pi(in)$ can, with high probability, create an output state $y$ that has vanishingly small probability under $\pi(in)$. The proposed explanation in earlier papers says that there was a sequence of intermediate stepping stones, each of which was likely to be generated from its ancestors, and which ends up with y. For simplicity, all the concepts introduced were formulated in terms of TMs with one-dimensional tapes containing bit
strings. However, ultimately we might we want to use the introduced concepts reformulated in terms of
other computationally universal systems, like the two-dimensional cellular automata that underpin many digital life systems.

Lastly, stochastic depth might be useful as a measure of lawfulness of physical laws. 
There is a common supposition that the sequence of events in our universe must have low Kolmogorov complexity, 
so that it ``counts as being lawful, as having been generated by mathematical laws, rather than 
just being a lawless, random sequence''. Note though that
running a program via fully homomorphic encryption (FHE) rather than running it directly does not change its Kolmogorov complexity. (Though
to run it via an FHE \textit{and then also decrypt the results} would result in a composite program with slightly larger Kolmogorov complexity, 
due to the length of the decryption key.) Yet as a \textit{practical matter}, it's effectively impossible to distinguish that sequence from IID noise without the key.

So we could have a sequence of events that
 \textit{appear} to be purely random, to us, as a practical matter (and so would not ``count as mathematical laws''), even though they have
low Kolmogorov complexity. In such a situation they have low Kolmogorov complexity, but \textit{we} cannot distinguish them from a sequence of events with high Kolmogorov complexity. 

Now one might argue that even if low Kolmogorov complexity of the sequence is not a sufficient condition
for it to be considered lawful by we human analysts of that sequence, it is still a \textit{necessary} condition
for that sequence to be considered lawful. However, Chaitin's incompleteness theorem says it is impossible (uncomputable)
to compute the Kolmogorov complexity of any sequence if that complexity
is greater than a specific small value. So if we take that very small value to (approximately) equal the threshold for lawfulness, 
we cannot prove that such a necessary condition is violated even if it is. We propose that stochastic depth or a related measure could be useful to measure the lawfulness of patterns that avoids the problem with using Kolmogorov complexity.

\section{Conclusion}\label{Conclusion}
In this paper, we have introduced a type of stochastic process defined by a PTM and a class of complexity measures, called stochastic depth, defined for these TMs. An SPTM is a well-defined stochastic process with a given prior and a conditional distribution given by the update functions of the specified PTM.  SPTMs address a fundamental incongruence in the application of TM theory to real-world systems: many of these systems are effectively random. Stochastic depth is a type of novel complexity measure that utilizes the Bayesian inverse of the forward conditional that defines a PTM. Measures of this type have a desirable property that most complexity measures lack -- not only do they take into account the lengths of the computational paths, like Levin complexity, but they take into account how non-invertible the individual steps of a TM trajectory are. This is essentially a measure of how much information is lost over a TMs trajectory. Stochastic depth also has interesting connections to thermodynamics, and future work involves formalizing and further investigating these connections.

We have outlined some important potential use cases and argued that SPTMs are the natural setting to study assembly complexity and its criticisms. Future work includes extensions of the SPTM idea to other computationally universal systems, in particular 2D cellular automata. This is connected to applying the SPTM formalism to studies of artificial life, which are often in the form of a cellular automata. A particular avenue of artificial life studies is the ``stepping stone" phenomenon described in Sec. \ref{Applications}. Future work also includes proving the computability properties of stochastic depth beyond the deterministic limit. Overall, STPMs provide a novel framework for the study of real, stochastic systems that compute, and it provides a Bayesian inverse for stochastic generalizations of many useful complexity measures.

\pagebreak

\section*{Acknowledgments} The authors would like to thank the Santa Fe Institute for helping to support this research. We would also like to thank Peter Stadler, Carlos Gershenson, Blaise Aguera y Arcas, Robert Obryk, and Rif Saurous for helpful discussions.

\appendix

\bibliographystyle{unsrt}
\bibliography{SPTMPaperV3.bib}

\end{document}